\documentclass{article}
\usepackage[utf8]{inputenc}
\usepackage{macros}

\usepackage{graphicx}
\graphicspath{ {./images/} }

\title{Reweighted Solutions for Weighted Low Rank Approximation}
\author{
David P. Woodruff \\ Carnegie Mellon University \\ \texttt{dwoodruf@cs.cmu.edu} \and 
Taisuke Yasuda \\ Carnegie Mellon University \\ \texttt{taisukey@cs.cmu.edu}
}

\begin{document}

\maketitle

\thispagestyle{empty}
\begin{abstract}
Weighted low rank approximation (WLRA) is an important yet computationally challenging primitive with applications ranging from statistical analysis, model compression, and signal processing. To cope with the NP-hardness of this problem, prior work considers heuristics, bicriteria, or fixed parameter tractable algorithms to solve this problem. In this work, we introduce a new relaxed solution to WLRA which outputs a matrix that is not necessarily low rank, but can be stored using very few parameters and gives provable approximation guarantees when the weight matrix has low rank. Our central idea is to use the weight matrix itself to reweight a low rank solution, which gives an extremely simple algorithm with remarkable empirical performance in applications to model compression and on synthetic datasets. Our algorithm also gives nearly optimal communication complexity bounds for a natural distributed problem associated with this problem, for which we show matching communication lower bounds. Together, our communication complexity bounds show that the rank of the weight matrix provably parameterizes the communication complexity of WLRA. We also obtain the first relative error guarantees for feature selection with a weighted objective.
\end{abstract}

\clearpage
\setcounter{page}{1}

\section{Introduction}

The approximation of matrices by matrices of lower rank has been, and continues to be, one of the most intensely studied and applied computational problems in statistics, machine learning, and signal processing. The classical approach to this problem is to approximate a matrix $\bfA\in\mathbb R^{n\times d}$ by a rank $k$ matrix $\tilde\bfA\in\mathbb R^{n\times d}$ that minimizes the Frobenius norm error
\[
    \norm{\bfA - \tilde\bfA}_F^2 \coloneqq \sum_{i=1}^n \sum_{j=1}^d \abs{\bfA_{i,j}-\tilde\bfA_{i,j}}^2,
\]
where $\rank(\tilde\bfA) \leq k$. This problem can be solved exactly by the singular value decomposition (SVD), which can be computed in polynomial time. We will write $\bfA_k$ to denote the optimal rank $k$ approximation to $\bfA$ in the Frobenius norm, and we will write $\bfA_{-k}\coloneqq \bfA - \bfA_k$ to denote the residual error of this approximation.

While this simple choice often gives satisfactory results, this loss function treats all entries of the matrix uniformly when trying to fit $\tilde\bfA$, which may not exactly align with the practitioner's desires if some of the entries are more crucial to fit than others. If one additionally has such information available in the form of non-negative weights $\bfW_{i,j}\geq 0$ that reflect some measure of importance of each of the entries $(i,j)$, then this can be encoded in the loss function by incorporating weights as
\[
    \norm{\bfA - \tilde\bfA}_{\bfW,F}^2 \coloneqq \sum_{i=1}^n \sum_{j=1}^d \bfW_{i,j}^2\cdot \abs{\bfA_{i,j}-\tilde\bfA_{i,j}}^2,
\]
where $\rank(\tilde\bfA) \leq k$. This problem is known as the \emph{weighted low rank approximation (WLRA)} problem. We write $\bfA\circ\bfB$ to denote the entrywise product for two matrices $\bfA$ and $\bfB$, so we may also write
\[
    \norm{\bfA-\tilde\bfA}_{\bfW,F}^2 \coloneqq \norm{\bfW\circ (\bfA-\tilde\bfA)}_F^2 = \norm{\bfW\circ \bfA-\bfW\circ\tilde\bfA}_F^2
\]

The incorporation of weights into the low rank approximation problem gives this computational problem an incredible versatility for use in a long history of applications starting with its use in factor analysis in the early statistical literature \cite{You1941}. A popular special case is the \emph{matrix completion} problem \cite{RS2005, CT2010, KMO2010}, where the weights $\bfW\in\{0,1\}^{n\times d}$ are binary and encode whether a given entry of $\bfA$ is observed or not. This primitive has been useful in the design of recommender systems \cite{KBV2009, CLZLS2015, LKLSB2016}, and has been famously applied in the 2006 Netflix Prize problem. More generally, the weights $\bfW$ can be used to reflect the variance or number of samples obtained for each of the entries, so that more ``uncertain'' entries can influence the objective function less \cite{AI2002, SJ2003}. In the past few years, weighted low rank approximation has also been used to improve model compression algorithms, especially those for large scale LLMs, based on low rank approximations of weight matrices by taking into account the importance of parameters \cite{ALLMR2016, HHCLSJ2022, HHWLSJ2022}. Given the rapid growth of large scale machine learning models, model compression techniques such as WLRA are expected to bring high value to engineering efforts for these models. Other applications of WLRA include ecology \cite{RJMS2019, KHWH2022}, background modeling \cite{LDR2017, DLR2018}, computational biology \cite{THS2022}, and signal processing \cite{Shp1990, LPW1997}.

Approximation algorithms have long been considered for efficient low rank approximation problems, and we formalize the approximation guarantee that we study in Definition \ref{def:wlra}.

\begin{Definition}[Approximate weighted low rank approximation]
\label{def:wlra}
Let $\bfW\in\mathbb R^{n\times d}$ be non-negative, let $\bfA\in\mathbb R^{n\times d}$, and let $k\in\mathbb N$. Then in the \emph{$\kappa$-approximate rank $k$ weighted low rank approximation} problem, we seek to output a matrix $\tilde\bfA\in\mathbb R^{n\times d}$ such that
\[
    \norm{\bfA-\tilde\bfA}_{\bfW,F} \leq \kappa\min_{\rank(\bfA')\leq k}\norm{\bfA-\bfA'}_{\bfW,F}.
\]
\end{Definition}

In Definition \ref{def:wlra}, we have purposefully under-specified requirements on $\tilde\bfA$. Part of this is to cope with the computational difficulty of WLRA. Indeed, while we ideally would like $\tilde\bfA$ to have rank at most $k$, solving for even an approximate such solution (with $\kappa = (1+1/\poly(n))$) is an NP-hard problem \cite{GG2011}. On the other hand, allowing for additional flexibility in the choice of $\tilde\bfA$ may still be useful as long as $\tilde\bfA$ satisfies some sort of ``parameter reduction'' guarantee. A common choice is to allow $\tilde\bfA$ to have rank $k' \geq k$ slightly larger than $k$, which is known as a \emph{bicriteria} guarantee. In this work, we will show a new relaxation of the constraints on $\tilde\bfA$ that allows us to achieve new approximation guarantees for WLRA.

\subsection{Our results}

We present our main contribution in Theorem \ref{thm:wlra}, which gives a simple approach to WLRA, under the assumption that the weight matrix $\bfW$ has low rank $\rank(\bfW)\leq r$. We note that this assumption is very natural and captures natural cases, for example when $\bfW$ has block structure, and has been motivated and studied in prior work \cite{RSW2016, BWZ2019}. We also empirically verify this assumption in our experiments. We defer a further discussion of the low rank $\bfW$ assumption to Section \ref{sec:intro-experiments} as well as prior works \cite{RSW2016, BWZ2019}.

The algorithm (shown in Algorithm \ref{alg:wlra}) that we propose is extremely simple: compute a rank $rk$ approximation of $\bfW\circ\bfA$, and then divide the result entrywise by $\bfW$. Note that if we exactly compute the low rank approximation step by an SVD, then the optimal rank $rk$ approximation $(\bfW\circ\bfA)_{rk}$ given by the SVD requires only $(n+d)rk$ parameters to store, and $\bfW$ also only requires $nr$ parameters to store. Thus, denoting the entrywise inverse of $\bfW$ by $\bfW^{\circ-1}$, the solution $\bfW^{\circ-1}\circ (\bfW\circ\bfA)_{rk}$ can be stored in a total of $O((n+d)rk)$ parameters, which is nearly optimal for constant rank $r = O(1)$.\footnote{If $\bfW_{i,j} = 0$, we take $\bfW_{i,j}^{\circ-1} = \infty$. Note that this entry is ignored by the cost, i.e., $(\bfW\circ \bfW^{\circ-1})_{i,j} = 0$ by convention.}

\begin{algorithm}
	\caption{Weighted low rank approximation}
	\textbf{input:} input matrix $\bfA\in\mathbb R^{n\times d}$, non-negative weights $\bfW\in\mathbb R^{n\times d}$ with rank $r$, rank parameter $k$. \\
	\textbf{output:} approximate solution $\tilde\bfA$. \\
	\begin{algorithmic}[1] %
        \STATE Compute a rank $rk$ approximation $\tilde\bfA_\bfW$ of $\bfW\circ\bfA$
        \STATE Return $\tilde\bfA \coloneqq \bfW^{\circ-1}\circ\tilde\bfA_\bfW$
	\end{algorithmic}\label{alg:wlra}
\end{algorithm}

While our discussion thus far has simply used the SVD to compute the rank $rk$ approximation $(\bfW\circ\bfA)_{rk}$, we obtain other useful guarantees by allowing for approximate solutions $\tilde\bfA_\bfW$ that only approximately minimize $\norm{\bfW\circ\bfA - \tilde\bfA_\bfW}_F$. For example, by computing the rank $rk$ approximation $\tilde\bfA_\bfW$ using faster randomized approximation algorithms for the SVD \cite{CW2013, MM2015, ACW2017}, we obtain algorithms for WLRA with similar running time. In general, we prove the following theorem:

\begin{restatable}{Theorem}{ThmWLRA}
\label{thm:wlra}
Let $\bfW\in\mathbb R^{n\times d}$ be a non-negative weight matrix with rank $r$. Let $\bfA\in\mathbb R^{n\times d}$ and let $k\in\mathbb N$. Suppose that $\tilde\bfA_\bfW\in\mathbb R^{n\times d}$ satisfies
\begin{align*}
    \norm{\bfW\circ\bfA - \tilde\bfA_\bfW}_F^2 &\leq \kappa \min_{\rank(\bfA') \leq rk}\norm{\bfW\circ\bfA - \bfA'}_F^2 \\
    &= \kappa \norm{(\bfW\circ\bfA)_{-rk}}_F^2
\end{align*}
and let $\tilde\bfA\coloneqq \bfW^{\circ-1}\circ\tilde\bfA_\bfW$, where $\bfW^{\circ-1}\in\mathbb R^{n\times d}$ denotes the entrywise inverse of $\bfW$. Then,
\[
    \norm{\bfA-\tilde\bfA}_{\bfW,F}^2 \leq \kappa \min_{\rank(\bfA')\leq k}\norm{\bfA-\bfA'}_{\bfW,F}^2
\]
In particular, we obtain a solution with $\kappa = (1+\eps)$ in running time $O(\nnz(\bfA)) + \tilde O(n(rk)^2/\eps + \poly(rk/\eps))$ by using randomized low rank approximation algorithms of \cite{ACW2017}.
\end{restatable}

We prove Theorem \ref{thm:wlra} in Section \ref{sec:wlra}. In the special case of binary weight matrices, our result shows that ``zero-filling'' the missing entries leads to relative error guarantees in this natural setting, which is perhaps surprising due to a number of works studying this algorithm that suggest otherwise \cite{BRN2010, WS2017}.

Note that as stated, the approximation given by Algorithm \ref{alg:wlra} may not always be desirable, since in general, $\bfW^{\circ -1}$ cannot be computed without multiplying out the low rank factors of $\bfW$. However, we show in Lemma \ref{lem:structured-entrywise-inverse} that for a broad family of structured matrices formed by the sum of support-disjoint rank 1 matrices and a sparse matrix, $\bfW^{\circ-1}$ can in fact be stored and applied in the same time as $\bfW$. These capture a large number of commonly used weight matrix patterns in practice, such as Low-Rank Plus Sparse, Low-Rank Plus Diagonal, Low-Rank Plus Block Diagonal, Monotone Missing Data Pattern, and Low-Rank Plus Banded matrices \cite{MMW2021} (see Corollary \ref{cor:structured-entrywise-inverse}). These results are proved in Appendix \ref{sec:structured-inverse}.

\begin{restatable}{Lemma}{StructuredInverse}
\label{lem:structured-entrywise-inverse}
Let $\bfW\in\mathbb R^{n\times d}$ be structured as $\bfW = \bfE + \sum_{i=1}^{r'} \bfS_i$, where $\bfS_i$ are rank 1 matrices with disjoint rectangular supports $S_i\times T_i$ for $S_i\subseteq[n]$ and $T_i\subseteq[d]$, and $\bfE$ is a sparse matrix with $\nnz(\bfE)$ non-zero entries. Let $\bfA = \bfU\bfV$ for $\bfU\in\mathbb R^{n\times k}$ and $\bfV\in\mathbb R^{k\times d}$ be a rank $k$ matrix. Then, $\bfW^{\circ -1}\circ \bfA$ can be stored in $O(\nnz(\bfE) + \sum_{i=1}^{r'} \abs{S_i} + \abs{T_i})$ space and can be applied to a vector $\bfx\in\mathbb R^d$ in $O(\nnz(\bfE) + \sum_{i=1}^{r'} \abs{S_i} + \abs{T_i})$ time. Furthermore, $\bfW$ has rank at most $r = \nnz(\bfE) + r'$.
\end{restatable}

\begin{restatable}{Corollary}{StructuredInverseCor}
\label{cor:structured-entrywise-inverse}
Let $\bfW\in\mathbb R^{n\times d}$. The following hold:
\begin{itemize}
    \item \textbf{Low-Rank Plus Sparse}: Suppose that $\bfW$ has at most $t$ non-zeros per row. Then, $\bfW^{\circ-1}$ can be stored and applied in $O(nt)$ time and space.
    \item \textbf{Low-Rank Plus Diagonal}: Suppose that $\bfW$ is all ones except for zeros along the diagonal. Then, $\bfW^{\circ-1}$ can be applied and stored in $O(n)$ time and space.
    \item \textbf{Low-Rank Plus Block Diagonal}: Suppose that $\bfW$ is all ones except for $r$ block diagonal matrices that are zeros. Then, $\bfW^{\circ -1}$ can be applied and stored in $O(nr)$ time and space.
    \item \textbf{Monotone Missing Data Pattern}: Suppose that $\bfW$ is a rank $r$ matrix where each row is a prefix of all ones followed by a suffix of all zeros. Then, $\bfW^{\circ -1}$ can be applied and stored in $O(nr)$ time and space.
    \item \textbf{Low-Rank Plus Banded}: Suppose that $\bfW$ is all ones except for zeros on ``band'' entries, i.e., $\bfW_{i,j} = 0$ for $\abs{i-j}\leq p$. Then, $\bfW^{\circ -1}$ can be applied and stored in $O(np)$ time and space.
\end{itemize}
\end{restatable}

Thus, our results yield efficient algorithms with provable approximation guarantees for a wide class of structured weight matrices encountered in practice. Furthermore, for general weight matrices, our results can be applied by first computing a low rank approximation of the weight matrices or an approximation by one of the above structured classes of weight matrices for further improvements in efficiency.

\subsubsection{Column subset selection for weighted low rank approximation}

Another advantage of allowing for approximation algorithms for computing low rank approximations to $\bfW\circ\bfA$ is that we can employ \emph{column subset selection} approaches to low rank approximation \cite{FKV2004, DV2006, DMM2006b, DMM2008, BWZ2016, ABFMRZ2016}. That is, it is known that the Frobenius norm low rank approximation problem admits $(1+\eps)$-approximate low rank approximations whose left factor is formed by a subset of at most $O(k/\eps)$ columns of the input matrix. In particular, these results show the existence of approximate solutions to the low rank approximation problem that preserve the sparsity of the input matrix, and thus can lead to a reduced solution size when the input matrix has sparse columns. Furthermore, column subset selection solutions to low rank approximation give a natural approach for \emph{unsupervised feature selection}. Thus, as a corollary of Theorem \ref{thm:wlra}, we obtain the first relative error guarantee for unsupervised feature selection with a weighted Frobenius norm objective. Weaker additive error guarantees were previously studied by \cite{Dai2023, AY2023}\footnote{The result of \cite{Dai2023} contained an error, which we correct, tighten, and simplify in Appendix \ref{sec:row-norm-sampling}.}.

\begin{Corollary}[Column subset selection for weighted low rank approximation]
\label{cor:wlra-css}
There is an algorithm that computes a subset $S\subseteq[d]$ of $\abs{S} = O(rk/\eps)$ columns and $\bfX\in\mathbb R^{\abs{S}\times d}$ such that
\begin{align*}
    &\norm*{\bfA - \bfW^{\circ-1}\circ ((\bfW\circ\bfA)\lvert^S\bfX)}_{\bfW,F}^2 \\
    \leq~& (1+\eps)\min_{\rank(\bfA')\leq k}\norm{\bfA-\bfA'}_{\bfW,F}^2
\end{align*}
where for a matrix $\bfB\in\mathbb R^{n\times d}$, $\bfB\vert^S$ denotes the matrix formed by the columns of $\bfB$ indexed by $S$.
\end{Corollary}
\begin{proof}
This follows from Theorem \ref{thm:wlra} by computing the rank $rk$ approximation $\tilde\bfA_\bfW$ to $\bfW\circ\bfA$ via column subset selection algorithms given by, e.g., \cite{BWZ2016}.
\end{proof}

Note that in Corollary \ref{cor:wlra-css}, the approximation $\bfW^{\circ-1}\circ ((\bfW\circ\bfA)\lvert^S\bfX)$ only depends on $\bfA$ through the columns $\bfA\lvert^S$, and thus giving an approach to column subset selection with a weighted objective.

\subsubsection{Nearly optimal communication complexity bounds}

As a consequence of Corollary \ref{cor:wlra-css}, we obtain another important result for WLRA in the setting of \emph{communication complexity}. Here, we obtain nearly optimal communication complexity bounds for constant factor approximations (i.e., $\kappa = O(1)$) to distributed WLRA for a wide range of parameters. While many works have studied distributed LRA in depth \cite{Sar2006, CW2009, CW2013, MBZ2010, KVW2014, GLPW2016, BWZ2016}, we are surprisingly the first to initiate a study of this problem for WLRA.

The communication setting we consider is as follows. We have two players, Alice and Bob, where Alice has an input matrix $\bfA$ and would like to communicate an approximate WLRA solution to Bob. Communication complexity is of great interest in modern computing, where exchanging bits can be a critical bottleneck in large scale computation. While we consider two players in this discussion for simplicity, our algorithms also apply to a distributed computing setting, where the columns of the input matrix are partitioned among $m$ servers as $m$ matrices $\bfA^{(1)}, \bfA^{(2)}, \dots, \bfA^{(m)}$, and some central coordinator outputs a WLRA of the concatenation $\bfA = [\bfA^{(1)}, \bfA^{(2)}, \dots, \bfA^{(m)}]$ of these columns.

\begin{Definition}[WLRA: communication game]
\label{def:wlra-comm}
Let Alice and Bob be two players. Let $\bfW\in\mathbb Z^{n\times d}$ be non-negative, let $\bfA\in\mathbb Z^{n\times d}$, and let $k\in\mathbb N$. Furthermore, let $\bfW$ and $\bfA$ have entries at most $\bfW_{i,j}, \abs{\bfA_{i,j}} \leq \poly(nd)$. We let both Alice and Bob receive the weight matrix $\bfW$ as input, and we give only Alice the input matrix $\bfA$. We say that Alice and Bob solve the \emph{$\kappa$-approximate rank $k$ weighted low rank approximation communication game} using $B$ bits of communication if Alice sends at most $B$ bits to Bob, and Bob outputs \emph{any} matrix $\tilde\bfA\in\mathbb R^{n\times d}$ satisfying
\[
    \norm{\bfA - \tilde\bfA}_{\bfW,F} \leq \kappa \min_{\rank(\bfA') \leq k} \norm{\bfA - \bfA'}_{\bfW,F}.
\]
\end{Definition}

Suppose that $\bfA$ has columns which each have at most $s$ non-zero entries. Then, the solution given by Corollary \ref{cor:wlra-css} can be communicated to Bob using just $O(srk/\eps + rkd)$ numbers ($O(srk/\eps)$ for the $O(rk/\eps)$ columns of $\bfA$ and $O(rkd)$ for $\bfX$), or $O((srk/\eps + rkd)\log(nd))$ bits under our bit complexity assumptions. Thus, when the number of columns $d$ is at most the column sparsity $s$, then we obtain an algorithm which uses only $O((srk/\eps)\log(nd))$ bits of communication. More generally, if the columns of $\bfA$ are distributed among $m$ servers, then a solution can be computed using $O((msrk/\eps)\log(nd))$ bits of communication by using work of \cite{BWZ2016}.

In fact, we show a nearly matching communication lower bound. In particular, we show that $\Omega(srk)$ bits of communication is required to output \emph{any} matrix (not necessarily structured) that achieves a weighted Frobenius norm loss that is \emph{any} finite factor within the optimal solution. Our lower bound is information-theoretic, and also immediately implies an $\Omega(msrk)$ bit lower bound in the distributed setting of $m$ servers if each server must output a solution, as considered by \cite{BWZ2016}.

\begin{restatable}{Theorem}{ThmCommLB}
\label{thm:comm-lb}
Let $\bfW$ be a binary block diagonal mask (Definition \ref{def:block-diag-mask}) and let $k\in\mathbb N$. Suppose that a randomized algorithm solves, for every $\bfC\in\mathbb Z^{n\times n}$ with at most $s$ non-zero entries in each column, the $\kappa$-approximate weighted low rank approximation problem on input $\bfC$ using $B$ bits of communication with probability at least $2/3$, for any $1 \leq \kappa < \infty$. If $s, k \leq n/r$, then $B = \Omega(srk)$.
\end{restatable}

By proving a nearly tight communication complexity bound of $\tilde\Theta(rsk)$ for computing constant factor WLRAs, we arrive at the following qualitative observation: \emph{the rank $r$ of the weight matrix $\bfW$ parameterizes the communication complexity of WLRA}. A similar conclusion was drawn for the \emph{computational} complexity of WLRA in the work of \cite{RSW2016}, where it was shown that WLRA is fixed parameter tractable in the parameter $r$, and also must have running time exponential in $r$ under natural complexity theoretic assumptions. Thus, an important contribution of our work is to provide further evidence, both empirical and theoretical, that the rank $r$ of the weight matrix $\bfW$ is a natural parameter to consider when studying WLRA.

\subsubsection{Experiments}
\label{sec:intro-experiments}

We demonstrate the empirical performance of our WLRA algorithms through experiments for model compression tasks. This application of WLRA was suggested by \cite{HHCLSJ2022, HHWLSJ2022}, which we find to be a particularly relevant application of weighted low rank approximation due to the trend of extremely large models. In the model compression setting, we wish to approximate the hidden layer weight matrices of neural networks by much smaller matrices. A classical way to do this is to use low rank approximation \cite{SKSAR2013, KPYCYS2015, CSLCH2018}. While this often gives reasonable results, the works of \cite{HHCLSJ2022, HHWLSJ2022} show that significant improvements can be obtained by taking into account the importance of each of the parameters in the LRA problem. We thus conduct our experiments in this setting.

We first show in Section \ref{sec:low-rank-w} that the importance matrices arising this application are indeed very low rank. We may interpret this phenomenon intuitively: we hypothesize that the importance score of some parameter $\bfA_{i,j}$ is essentially the product of the importance of the corresponding input $i$ and the importance of the corresponding output $j$. This observation may be of independent interest, and also empirically justifies the low rank weight matrix assumption that we make in this work, as well as works of \cite{RSW2016, BWZ2019}. While WLRA with a rank 1 weight matrix is known to be solvable efficiently via the SVD, our result shows that general low rank weight matrices also yield efficient algorithms via the SVD.

Next in Section \ref{sec:approx-eff}, we conduct experiments which demonstrate the superiority of our methods in practice. Of the algorithms that we compare to, an expectation-minimization (EM) approach of \cite{SJ2003} gives the smallest loss albeit with a very high running time, and our algorithm nearly matches this loss with an order of magnitude lower running time. We also show that this solution can be refined with EM, producing the best trade-off between efficiency and accuracy. One of the baselines we compare is a sampling algorithm of \cite{Dai2023}, whose analysis contains an error which we correct, simplify, and tighten.

\subsection{Related work}

We survey a number of related works on approximation algorithms for weighted low rank approximation. One of the earliest algorithms for this problem is a natural EM approach proposed by \cite{SJ2003}. Another related approach is to parameterize the low rank approximation $\tilde\bfA$ as the product $\bfU\bfV$ of two matrices $\bfU\in\mathbb R^{n\times k}$ and $\bfV\in\mathbb R^{k\times d}$ and alternately minimize the two matrices, known as \emph{alternating least squares}. This algorithm has been studied in a number of works \cite{HMLZ2015, LLR2016, SYYZ2023}. The work of \cite{BRW2021} proposes an approach to weighted low rank approximation based on a greedy pursuit, where rank one factors are iteratively added based on an SVD of the gradient matrix. Finally, fixed parameter tractable algorithms have been considered in \cite{RSW2016, BWZ2019} based on sketching techniques.

\section{Approximation algorithms}
\label{sec:wlra}

The following simple observation is the key idea behind Theorem \ref{thm:wlra}:

\begin{Lemma}
\label{lem:entrywise-product-rank}
Let $\bfW,\bfA'\in\mathbb R^{n\times d}$ with $\rank(\bfW)\leq r$ and $\rank(\bfA')\leq k$. Then, $\rank(\bfW\circ\bfA') \leq rk$.
\end{Lemma}
\begin{proof}
Since $\rank(\bfW)\leq r$, it can be written as $\bfW = \sum_{i=1}^r \bfu_i \bfv_i^\top$ for $\bfu_i\in\mathbb R^n$ and $\bfv_i\in\mathbb R^d$. Then,
\[
    \bfW\circ\bfA' = \sum_{i=1}^r (\bfu_i\bfv_i^\top)\circ \bfA' = \sum_{i=1}^r \diag(\bfu_i) \bfA' \diag(\bfv_i)
\]
so $\bfW\circ\bfA'$ is the sum of $r$ matrices, each of which is rank $k$. Thus, $\bfW\circ\bfA'$ has rank at most $rk$.
\end{proof}

Using Lemma \ref{lem:entrywise-product-rank}, we obtain the following:

\ThmWLRA*
\begin{proof}
Note that $\norm{\bfW^{\circ -1}\circ \tilde\bfA_\bfW - \bfA}_{\bfW,F}^2 = \norm{\tilde\bfA_\bfW - \bfW\circ \bfA}_F^2$, which is at most $\kappa\norm{(\bfW\circ\bfA)_{-rk}}_F^2$ by assumption. On the other hand for any rank $k$ matrix $\bfA'$, $\norm{\bfA' - \bfA}_{\bfW, F} = \norm{\bfW\circ \bfA' - \bfW\circ \bfA}_F$ can be lower bounded by $\norm{(\bfW\circ \bfA)_{-rk}}_F$ since $\bfW\circ \bfA'$ has rank at most $rk$ by Lemma \ref{lem:entrywise-product-rank}. Thus,
\begin{align*}
    \norm{\bfW^{\circ -1}\circ \tilde\bfA_\bfW - \bfA}_{\bfW,F}^2 &\leq \kappa \norm{(\bfW\circ\bfA)_{-rk}}_F^2 \\
    &\leq \kappa \min_{\rank(\bfA')\leq k} \norm{\bfA-\bfA'}_{\bfW,F}^2.\qedhere
\end{align*}
\end{proof}

\section{Matrices with structured entrywise inverses}
\label{sec:structured-inverse}

We present a general lemma which shows how to handle a wide family of structured matrices which often arise in practice as weight matrices for weighted low rank approximation.

\StructuredInverse*
\begin{proof}
Let $\bfS = \sum_{i=1}^r\bfS_i$. We can then write $\bfW^{\circ-1} = \bfE' + \bfS^{\circ-1}$, where $\bfE'$ is a sparse matrix with $\nnz(\bfE') = \nnz(\bfE)$. Since $\bfS_i$ have disjoint supports, we can also write $\bfS^{\circ-1} = \sum_{i=1}^r \bfS_i^{\circ -1}$. Thus,
\begin{align*}
    (\bfW^{\circ-1}\circ\bfA)\bfx &= \parens*{\bfE' + \parens*{\sum_{i=1}^r \bfS_i^{\circ-1}}\circ \bfA}\bfx \\
    &= (\bfE'\circ \bfA)\bfx + \sum_{i=1}^r (\bfS_i^{\circ-1}\circ\bfA)\bfx
\end{align*}
Note that $\nnz(\bfE'\circ\bfA) \leq \nnz(\bfE)$, so this can be stored and applied in $O(\nnz(\bfE))$ time and space. For each $i$, $\bfS_i^{\circ-1}$ is just a rank 1 matrix supported on $S_i\times T_i$, so this can be stored and applied in $O(\abs{S_i}+\abs{T_i})$ time and space.
\end{proof}

As a corollary of Lemma \ref{lem:structured-entrywise-inverse}, we show that we can efficiently handle all five families of commonly encountered weight matrices discussed in \cite{MMW2021}. We refer to \cite{MMW2021} on the large body of work studying these classes of weight matrices.

\StructuredInverseCor*
\begin{proof}
For the Low-Rank Plus Sparse weight matrices, $\bfW$ itself has at most $O(nt)$ non-zero entries and thus can be set to $\bfE$ in Lemma \ref{lem:structured-entrywise-inverse}. For the Low-Rank Plus Diagonal weight matrices, $\bfW$ can be written as the sum of the rank 1 matrix of all ones and a sparse matrix supported on the diagonal. For the Low-Rank Plus Block Diagonal weight matrices, we can write the complement of the blocks along their rows as disjoint rank 1 matrices. Since there are at most $r$ blocks, this is the sum of at most $r+1$ disjoint rank 1 matrices. For the Monotone Missing Data Pattern weight matrices, there can only be at most $r$ different patterns of rows, and these can be written as the sum of $r$ disjoint rank $1$ matrices. For the Low-Rank Plus Banded weight matrices, $\bfW$ can be written as the sum of the rank 1 matrix of all ones and a sparse matrix supported on the diagonal band.
\end{proof}

\section{Communication complexity bounds}
\label{sec:comm-comp}

We show that our approach to weighted low rank approximation in Theorem \ref{thm:wlra} gives nearly optimal bounds for this problem in the setting of communication complexity.

Our first result is an upper bound for the communication game in Definition \ref{def:wlra-comm}.

\begin{Theorem}
Let $\bfW\in\mathbb Z^{n\times d}$ be a non-negative rank $k$ weight matrix and let $\bfA\in\mathbb Z^{n\times d}$ be an input matrix with at most $s$ non-zero entries in each column. There is an algorithm which solves the $(1+\eps)$-approximate weighted low rank approximation communication game (Definition \ref{def:wlra-comm}) using at most $B = O((srk/\eps + rkd)\log(nd))$ bits of communication.
\end{Theorem}
\begin{proof}
The algorithm is to use the column subset selection-based WLRA algorithm of Corollary \ref{cor:wlra-css} and then to send the columns of $\bfA$ indexed by the subset $S$ and $\bfX$.
\end{proof}

On the other hand, we show a communication complexity lower bound showing that the number of bits $B$ exchanged by Alice and Bob must be at least $\Omega(rsk)$. Our lower bound holds even when the weight matrix $\bfW$ is the following simple binary matrix.

\begin{Definition}[Block diagonal mask]
\label{def:block-diag-mask}
Let $r\in\mathbb N$ and let $n$ be an integer multiple of $r$. Then, $\bfW\in\{0,1\}^{n\times n}$ is the \emph{block diagonal mask} associated with these parameters if $\bfW$ is the $r\times r$ block diagonal matrix with diagonal blocks given by the $n/r\times n/r$ all ones matrix and off-diagonal blocks given by the $n/r \times n/r$ all zeros matrix.
\end{Definition}

We give our communication complexity lower bound in the following theorem.

\ThmCommLB*
\begin{proof}
Let $\bfA_{\mathrm{dense}}\in\{0,1\}^{sr \times k}$ be a uniformly random binary matrix, and let $\bfA_{\mathrm{pad}}\in\{0,1\}^{n\times n/r}$ be formed by padding the columns of $\bfA_{\mathrm{dense}}$ with $n/r - k$ zero columns and padding each block of $s$ contiguous rows with $n/r - s$ zero rows. For $j\in[r]$, let $\bfA_{\mathrm{pad}}^{(j)}$ be the restriction of $\bfA_{\mathrm{pad}}$ to the $j$th contiguous block of $n/r$ rows. We then construct $\bfA\in\mathbb R^{n\times n}$ by horizontally concatenating $r$ copies of $\bfA_{\mathrm{pad}}$.

Note that the optimal rank $k$ approximation achieves $0$ loss in the $\bfW$-weighted Frobenius norm. Indeed, we can take $\bfA^*$ to be the horizontal concatenation of $r$ copies of $\bfA_{\mathrm{pad}}$. Since $\bfA_{\mathrm{pad}}$ has rank $k$, $\bfA^*$ also has rank $k$. Furthermore, on the $j$-th nonzero blocks of $\bfW$, $\bfA_{\mathrm{pad}}$ has the same entries as $\bfA_{\mathrm{pad}}^{(j)}$. Thus, an approximation $\tilde\bfA$ that achieves \emph{any} finite approximation factor $\kappa$ must exactly recover $\bfA$, restricted to the support of $\bfW$. In turn, this means that such an approximation $\tilde\bfA$ can also be used to recover $\bfA_{\mathrm{dense}}$.

It now follows by a standard information-theoretic argument that $B = \Omega(srk)$ (see Appendix \ref{app:comm-comp} for further details). 
\end{proof}

\section{Experiments}

As discussed in Section \ref{sec:intro-experiments}, we first conduct experiments for WLRA in the setting of model compression (as proposed by \cite{HHCLSJ2022, HHWLSJ2022}). In our experiments, we train a basic multilayer perceptron (MLP) on four image datasets, \texttt{mnist}, \texttt{fashion\_mnist}, \texttt{smallnorb}, and \texttt{colorectal\_histology} which were selected from the \texttt{tensorflow\_datasets} catalogue for simplicity of processing (e.g., fixed feature size, no need for embeddings, etc.). We then compute a matrix of importances of each of the parameters in a hidden layer of the MLP given by the Fisher information matrix. Finally, we compute a WLRA of the hidden layer matrix using the Fisher information matrix as the weights $\bfW$.

Our experiments are conducted on a 2019 MacBook Pro with a 2.6 GHz 6-Core Intel Core i7 processor. All code used in the experiments are available in the supplementary.

\begin{table}[ht]
\label{tab:datasets}
\caption{Datasets used in experiments}
\centering
\resizebox{\columnwidth}{!}{%
\begin{tabular}{ c c c c c }
\toprule
Dataset & Image dim. & Flattened dim. & Neurons & Matrix dim. \\
\hline
\texttt{mnist} & $(28, 28, 1)$ & $784$ & $128$ & $784 \times 128$ \\
\texttt{fashion\_mnist} & $(28, 28, 1)$ & $784$ & $128$ & $784 \times 128$ \\
\texttt{smallnorb} & $(96, 96, 1)$ & $9216$ & $1024$ & $9216 \times 1024$ \\
\texttt{colorectal\_histology} & $(150, 150, 3)$ & $67500$ & $1024$ & $67500 \times 1024$ \\
\bottomrule
\end{tabular}
}
\end{table}

\subsection{The low rank weight matrix assumption in practice}
\label{sec:low-rank-w}

We first demonstrate that for the task of model compression, the weight matrix is approximately low rank in practice. The weight matrix $\bfW$ in this setting is the empirical Fisher information matrix of the hidden layer weights $\bfA$, where the empirical Fisher information of the $(i,j)$-th entry $\bfA_{i,j}$ is given by
\[
    \bfW_{i,j} \coloneqq \mathbb E_{\bfx\sim\mathcal D}\bracks*{\parens*{\frac{\partial}{\partial\bfA_{i,j}}\mathcal L(\bfx;\bfA)}^2}
\]
where $\mathcal L(\bfx;\bfA)$ denotes the loss of the neural network on the data point $\bfx$ and hidden layer weights $\bfA$, and $\mathcal D$ denotes the empirical distribution (that is, the uniform distribution over the training data).

\begin{figure}[ht]%
    \centering
    \subfloat[\centering \texttt{mnist}]{{\includegraphics[width=4cm]{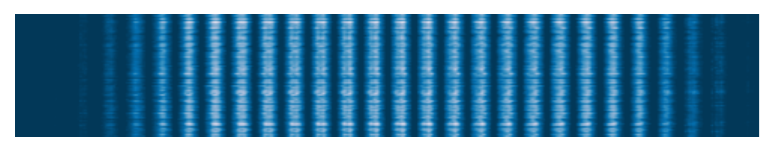} }}%
    \qquad
    \subfloat[\centering \texttt{fashion\_mnist}]{{\includegraphics[width=4cm]{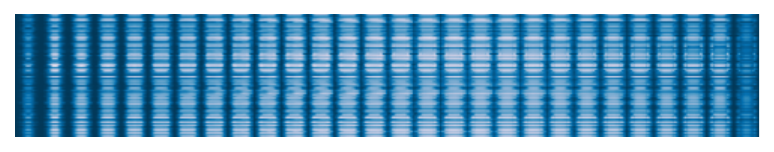} }}%
    \qquad
    \subfloat[\centering \texttt{smallnorb}]{{\includegraphics[width=4cm]{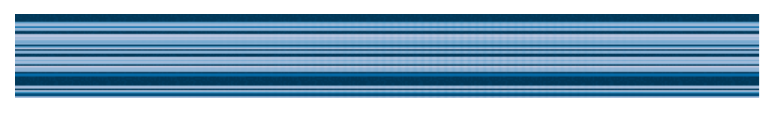} }}%
    \caption{Low rank structure of Fisher weight matrices}%
    \label{fig:fisher}%
\end{figure}

Plots of the empirical Fisher matrix (Figure \ref{fig:fisher}) reveal low rank structure to the matrices, and the spectrum of the Fisher matrix confirms that the vast majority of the Frobenius norm is contained in the first singular value (Table \ref{tab:spectrum}). We also plot the spectrum itself in Figure \ref{fig:spectrum} in the appendix.

\begin{table}[ht]
\caption{\% mass of Fisher matrix in 1st singular value}
\centering
\resizebox{0.5\columnwidth}{!}{%
\begin{tabular}{ c c c c c }
\toprule
Dataset & \% mass \\
\hline
\texttt{mnist} & 95.4\% \\
\texttt{fashion\_mnist} & 95.9\% \\
\texttt{smallnorb} & 99.9\% \\
\texttt{colorectal\_histology} & 99.3\% \\
\bottomrule
\end{tabular}
}
\label{tab:spectrum}
\end{table}

\subsection{Approximation quality and running time}
\label{sec:approx-eff}

In this section, we compare the performance of our Algorithm \ref{alg:wlra} (denoted as \texttt{svd\_w} in the following discussion) with a variety of previously proposed algorithms for WLRA. 

We consider the following algorithms: \texttt{adam}, \texttt{em}, \texttt{greedy}, \texttt{sample}, and \texttt{svd}, which we next explain in detail. We first consider \texttt{adam}, in which we simply parameterize the WLRA problem as an optimization problem in the factorized representation $\bfU\bfV$ for factors $\bfU\in\mathbb R^{n\times k}$ and $\bfV\in\mathbb R^{k\times d}$ \cite{BM2003}, and optimize this loss function using the Adam optimizer provided in the \texttt{tensorflow} library. Such an approach is well-studied for the standard low rank approximation problem \cite{LMZ2018, YD2021}, and empirically performs well for weighted low rank approximation as well. This was run for 100 epochs, with an initial learning rate of 1.0 decayed by a factor of 0.7 every 10 steps. The \texttt{em} algorithm was proposed by \cite{SJ2003} for the WLRA problem, and involves iteratively ``filling in'' missing values and recomputing a low rank approximation. In the experiments, we run 25 iterations. The \texttt{greedy} algorithm is a greedy basis pursuit algorithm proposed by \cite{BRW2021} and iteratively adds new directions to the low rank approximation by taking an SVD of the gradient of the objective. Similar algorithms were also studied in \cite{SGS2011, KEDGN2017, AS2021} for general rank-constrained convex optimization problems. The \texttt{sample} algorithm is a row norm sampling approach studied by \cite{Dai2023}. Finally, \texttt{svd} simply computes an SVD of the original matrix $\bfW$, without regard to the weights $\bfW$.

\begin{figure*}%
    \centering
    \includegraphics[width=11.5cm]{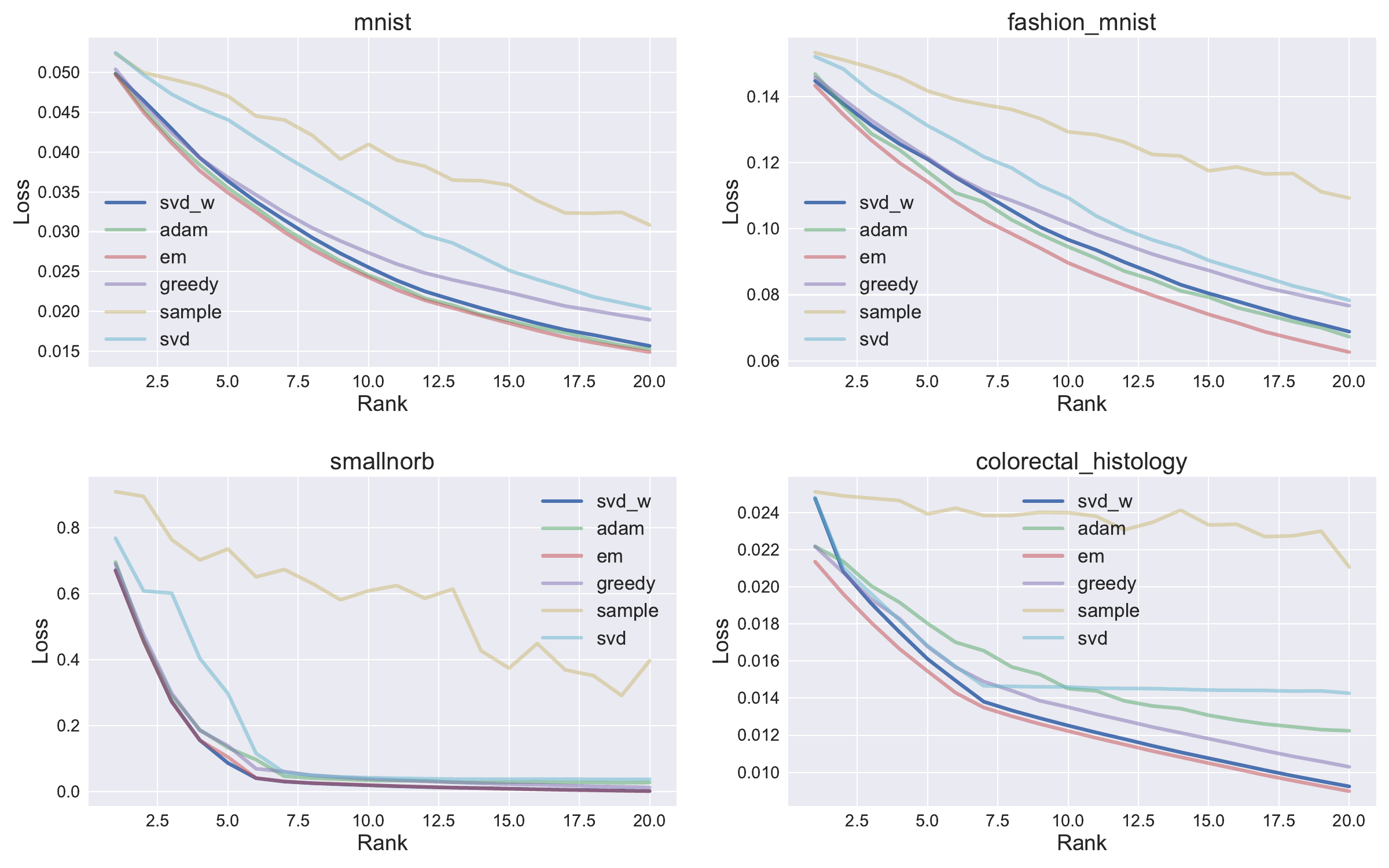}
        \caption{Fisher-weighted low rank approximation loss of weighted low rank approximation algorithms for model compression of four datasets. Results are averaged over $5$ trials.}%
    \label{fig:experiments-loss}%
\end{figure*}

\begin{figure*}%
    \centering
    \includegraphics[width=11.5cm]{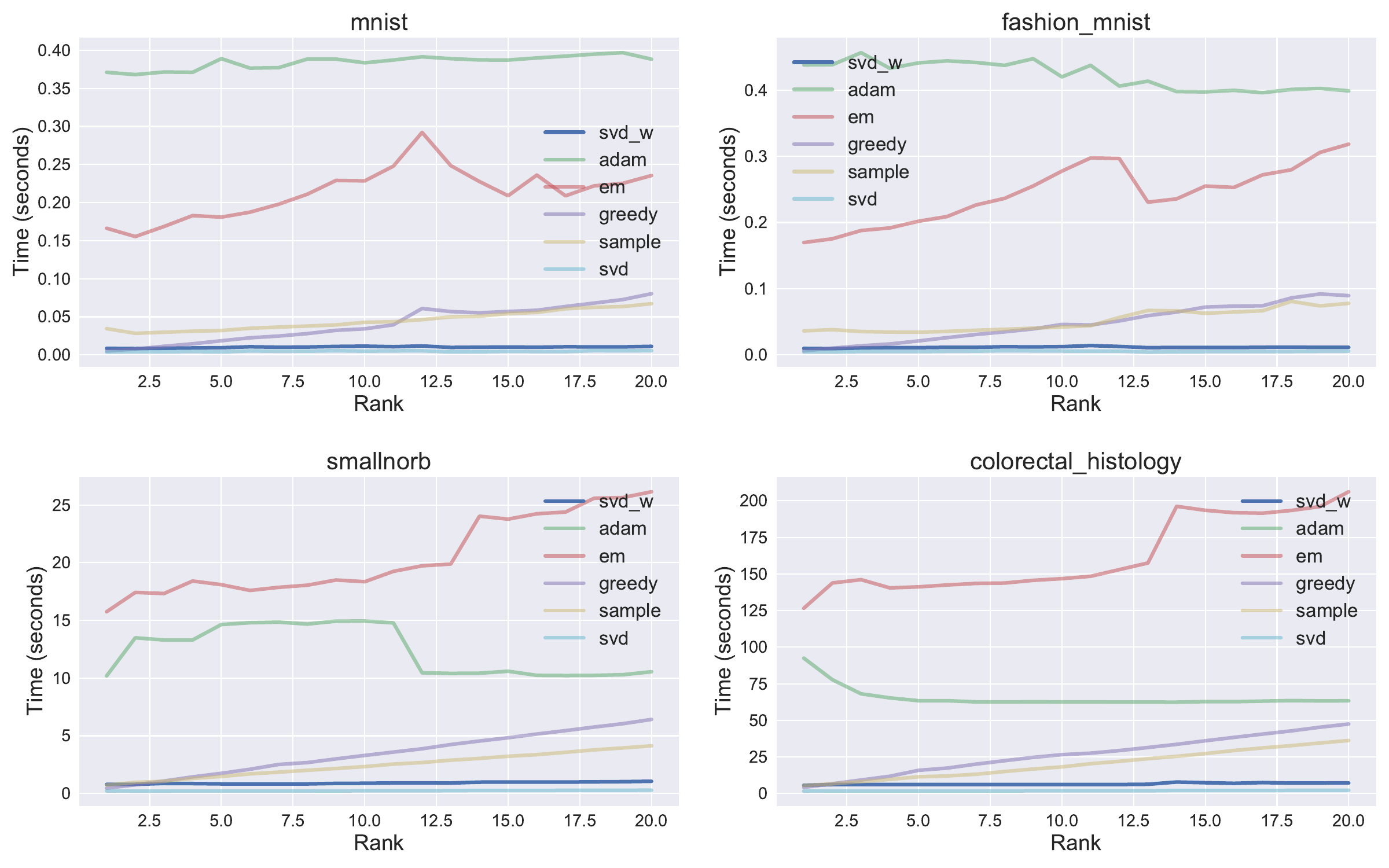}
    \caption{Running time of weighted low rank approximation algorithms for model compression of four datasets. Results are averaged over $5$ trials.}%
    \label{fig:experiments-time}%
\end{figure*}

We compute low rank approximations for ranks $1$ through $20$ on four datasets, and plot the loss and the running time against the rank in Figures \ref{fig:experiments-loss} and \ref{fig:experiments-time}, respectively. The values in the figures are tabulated at ranks $20$, $10$, and $5$ in Tables \ref{tab:experiments-20}, \ref{tab:experiments-10}, and \ref{tab:experiments-5} in the supplementary. We observe that our \texttt{svd\_w} algorithm performs among the best in the approximation loss (Figure \ref{fig:experiments-loss}), nearly matching the approximation quality achieved by much more computational expensive algorithms such as \texttt{adam} and \texttt{em}, while requiring much less computational time (Figure \ref{fig:experiments-time}).

\begin{figure*}%
    \centering
    \includegraphics[width=14cm]{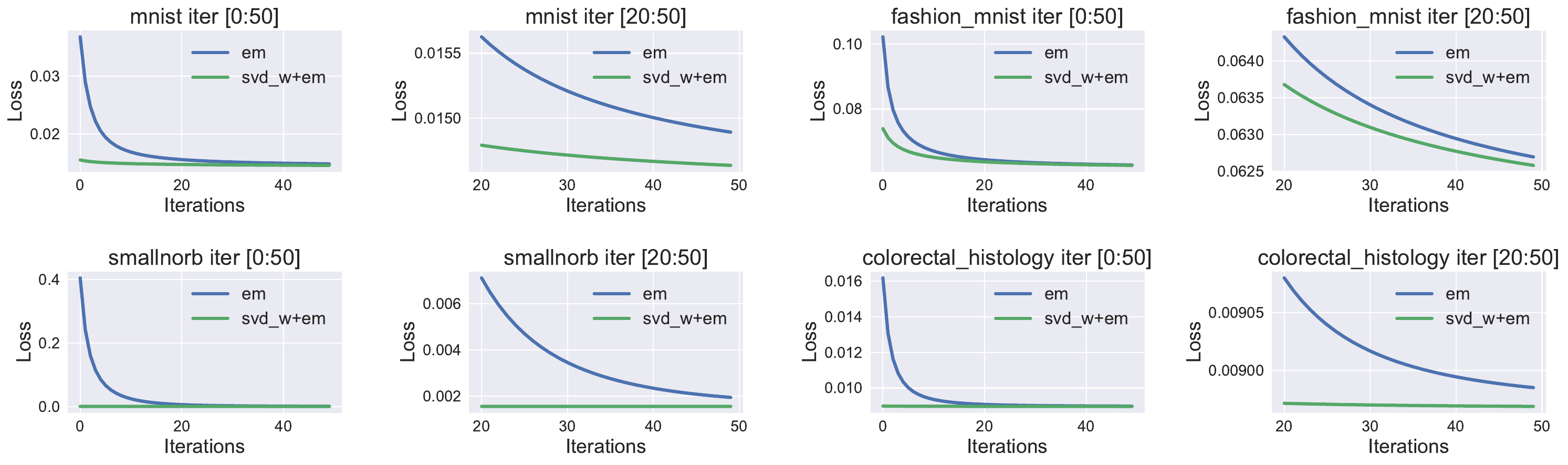}
    \caption{Improving the \texttt{svd\_w} solution with \texttt{em} iterations for a rank $20$ approximation.}%
    \label{fig:seed-em}%
\end{figure*}

While in some cases the \texttt{em} algorithm may eventually produce a better solution, we note that our \texttt{svd\_w} may be improved by initializing the \texttt{em} algorithm with this solution, which produces an algorithm which quickly produces a superior solution with many fewer iterations (Figure \ref{fig:seed-em}).

\subsection{Experiments on synthetic datasets}

\begin{figure}[H]%
    \centering
    \includegraphics[width=\columnwidth]{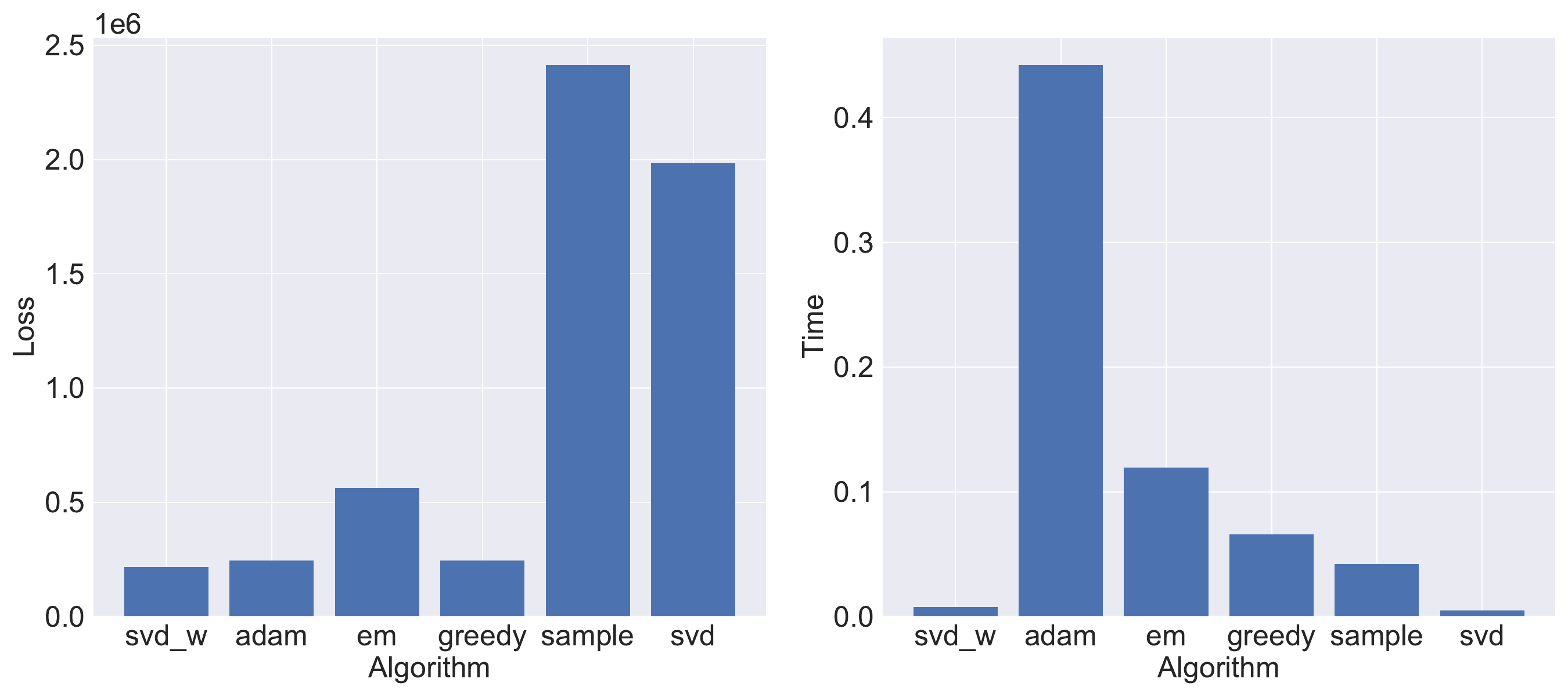}
    \caption{Loss and running time of WLRA on a synthetic dataset based on a mixture of Gaussians. Results are averaged over $5$ trials.}%
    \label{fig:mog}%
\end{figure}

Finally, we perform additional experiments on a synthetic datasets based on a mixture of Gaussians. In this experiment, we consider a uniform mixture of $k$ Gaussians in $d$ dimensions with diagonal covariance matrices, each which has variances that take at most $r$ distinct values. The input matrix $\bfA$ is taken to be $n$ i.i.d. observations of this distribution, while the weight matrix scales a Gaussian with variance $\sigma^2$ by $1/\sigma^2$. It can easily be observed that this weight matrix has rank at most $kr$, and thus our results apply. The variances for are chosen randomly by taking the $4$th powers of random Gaussian variables. The $4$th power is taken to make the variances more varied, which makes the WLRA problem more interesting. In Figure \ref{fig:mog}, we show the results for $n = 1000$, $d = 50$, $k = 5$, and $r = 3$. Our results again show that our algorithm achieves superior losses with a running time that is competitive with a standard SVD.

\section{Conclusion}

In this work, we studied new algorithms for the weighted low rank approximation problem, which has countless applications in statistics, machine learning, and signal processing. We propose an approach based on reweighting a low rank matrix, which is a novel class of relaxed solutions to the WLRA problem, and give provable guarantees under the assumption that the weight matrix $\bfW$ has low rank. Theoretically, this allows us to obtain an algorithm for WLRA with nearly optimal communication complexity, for which we show nearly matching communication complexity lower bounds, which shows that the rank of the weight matrix tightly parameterizes the communication complexity of this problem. We also give the first guarantees for column subset selection for weighted low rank approximation, which gives a notion of feature selection with a weighted objective. Finally, we show that in practice, our approach gives a highly efficient algorithm that outperforms prior algorithms for WLRA, particularly when combined with refinement using expectation-maximization.

\section*{Acknowledgements}

We thank the anonymous reviewers for useful feedback on improving the presentation of this work. We also thank Yucheng Dai for helpful discussions. David P.\ Woodruff and Taisuke Yasuda were supported by a Simons Investigator Award.

\bibliographystyle{alpha}
\bibliography{citations}

\appendix

\section{Missing proofs from Section \ref{sec:comm-comp}}
\label{app:comm-comp}

We provide the standard information-theoretic argument omitted in the proof of Theorem \ref{thm:comm-lb} in Section \ref{sec:comm-comp}.

Let $M$ denote the communication transcript between Alice and Bob, and let $\tilde\bfA_{\mathrm{dense}}$ denote the reconstruction of $\bfA_{\mathrm{dense}}$ based on the approximate weighted low rank approximation solution $\tilde\bfA$, which can be constructed based on $M$. Recall that the algorithm succeeds in outputting a correct approximation with probability at least $2/3$, so $\tilde\bfA_{\mathrm{dense}} = \bfA_{\mathrm{dense}}$ with probability at least $2/3$. Then by Fano's inequality (Theorem 2.10.1 of \cite{CT2006}), we have that
\begin{equation}
\label{eq:fano}
    H(\bfA_{\mathrm{dense}}\mid \tilde\bfA_{\mathrm{dense}}) \leq h(1/3) + \frac13 \log_2\abs{\mathcal A} = h(1/3) + \frac13 srk
\end{equation}
where $\mathcal A$ denotes the support of the random variable $\bfA_{\mathrm{dense}}$ and $h$ denotes the binary entropy function. It then follows from the data processing inequality (Theorem 2.8.1 of \cite{CT2006}) and the previous bound that the message length $B = \abs{M}$ is at least
\begin{align*}
    B &= \abs{M} \geq H(M) \geq I(M; \bfA_{\mathrm{dense}}) \\
    &\geq I(\tilde\bfA_{\mathrm{dense}}; \bfA_{\mathrm{dense}}) && \text{data processing inequality} \\
    &= H(\bfA_{\mathrm{dense}}) - H(\bfA_{\mathrm{dense}}\mid\tilde\bfA_{\mathrm{dense}}) \\
    &\geq rsk - (h(1/3) + srk/3) = \Omega(rsk) && \text{\eqref{eq:fano}}.\qedhere
\end{align*}

\section{Additional figures for experiments}

\begin{figure}[ht]%
    \centering
    \subfloat[\centering \texttt{mnist}]{{\includegraphics[width=6cm]{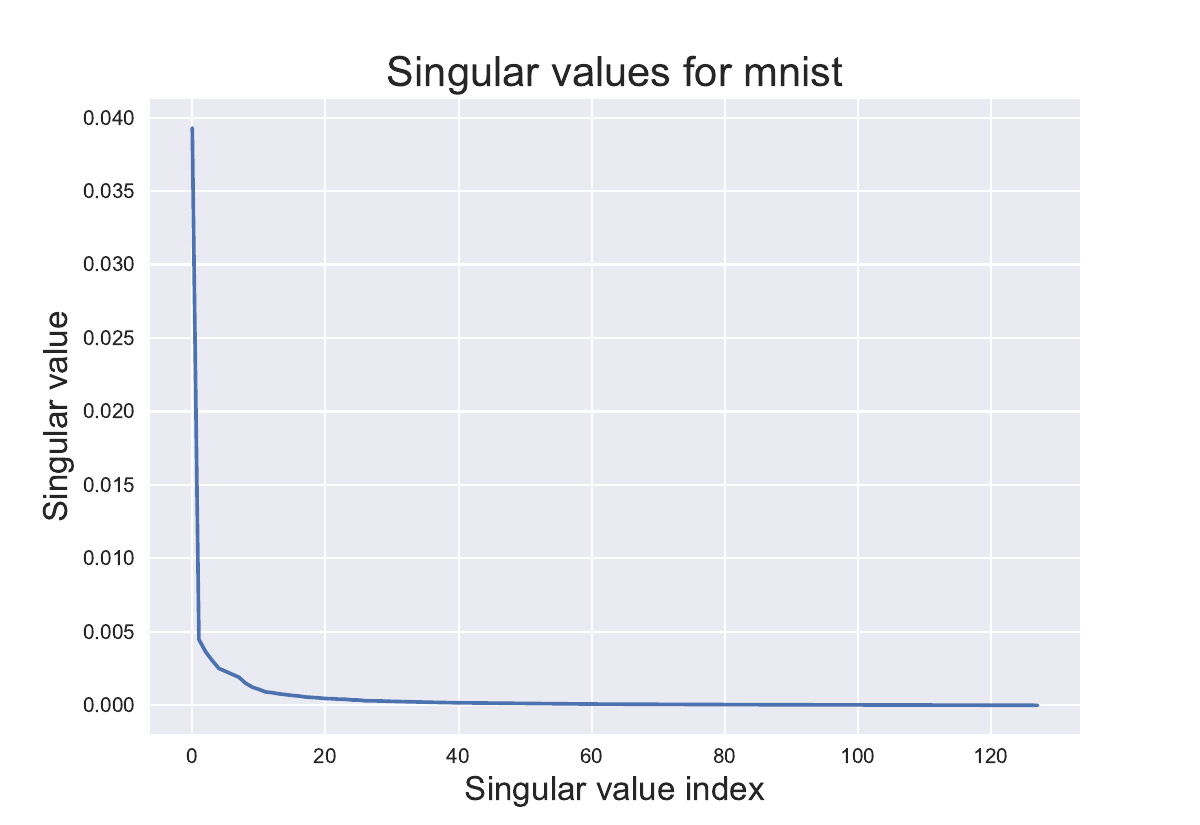} }}%
    \qquad
    \subfloat[\centering \texttt{fashion\_mnist}]{{\includegraphics[width=6cm]{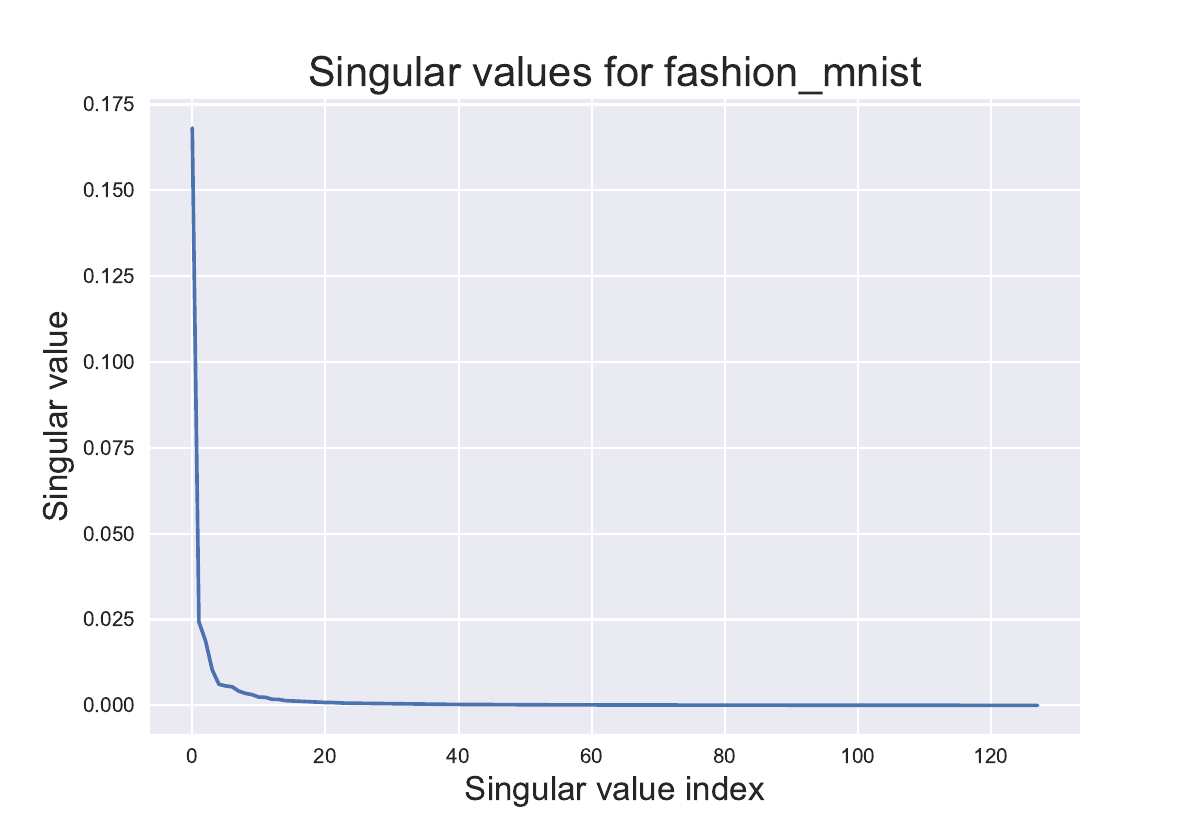} }}%
    \qquad
    \subfloat[\centering \texttt{smallnorb}]{{\includegraphics[width=6cm]{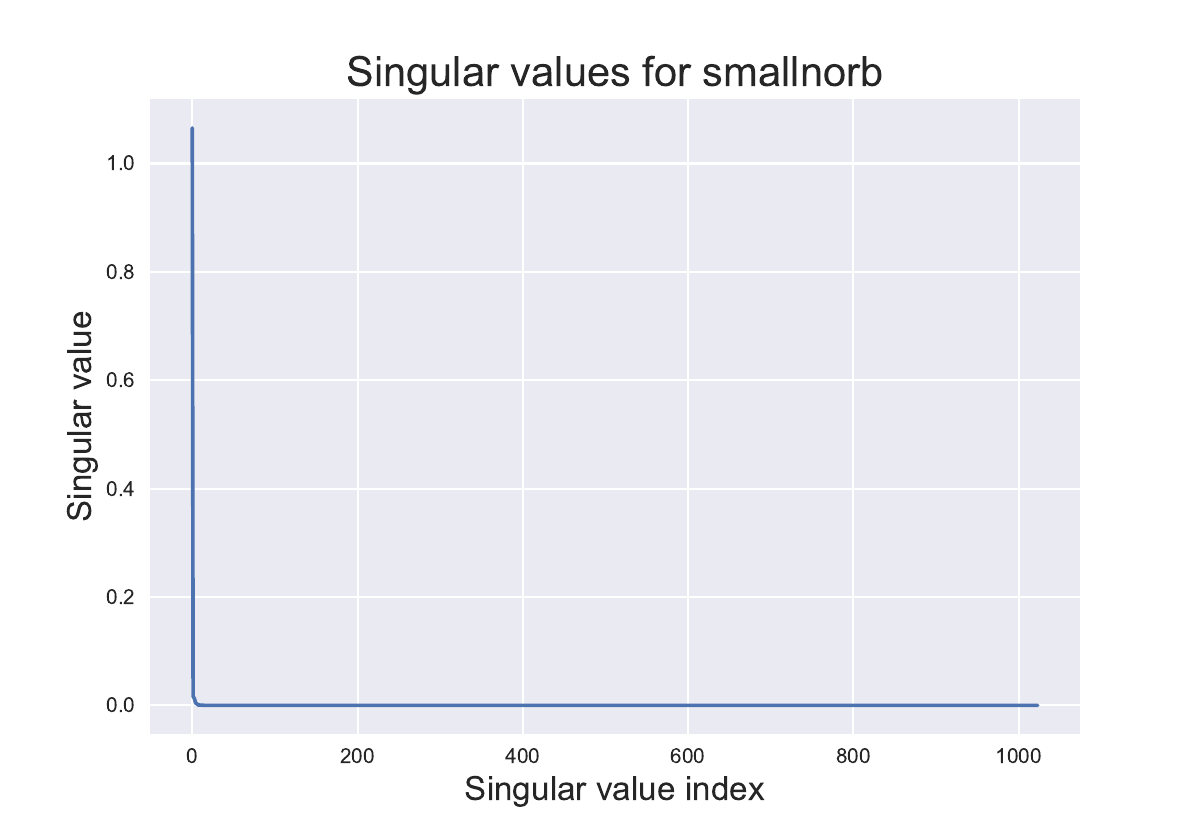} }}%
    \qquad
    \subfloat[\centering \texttt{colorectal\_histology}]{{\includegraphics[width=6cm]{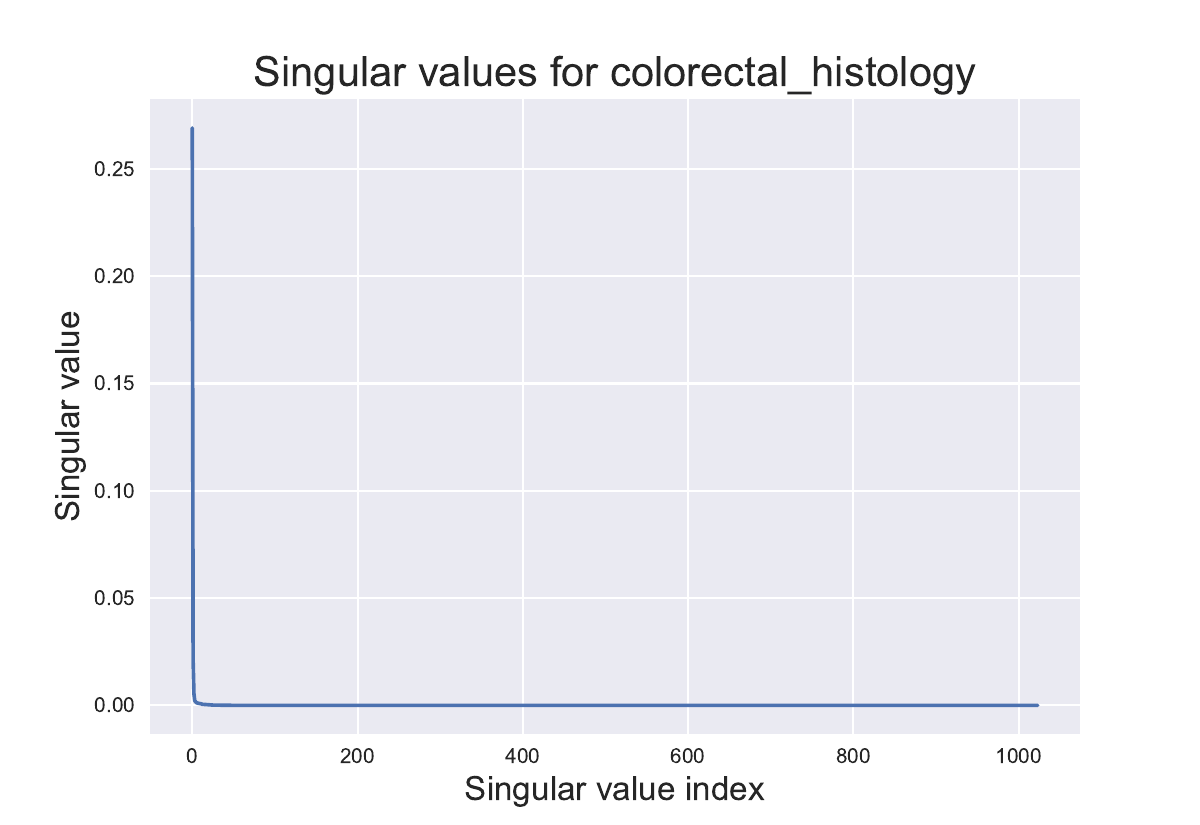} }}%
    \caption{Spectrum of Fisher weight matrices}%
    \label{fig:spectrum}%
\end{figure}

In Figure \ref{fig:spectrum}, we plot the spectrum of the weight matrices that we consider, showing that the assumption of a low rank weight matrix is a highly practical one.

In Tables \ref{tab:experiments-20} through \ref{tab:experiments-5}, we tabulate the approximation loss and running time of various algorithms for ranks $20$, $10$, and $5$, respectively.

\begin{table}[ht]
\caption{Fisher-weighted rank $20$ approximation loss and running time}
\label{tab:experiments-20}
\centering
\begin{tabular}{ l c c c c }
\toprule
Algorithm & \texttt{mnist} & \texttt{fashion\_mnist} & \texttt{smallnorb} & \texttt{colorectal\_histology} \\
\hline
\texttt{svd\_w} loss & 0.0157 & 0.0689 & 0.0023 & 0.0092 \\
\texttt{adam} loss & 0.0153 & 0.0673 & 0.0281 & 0.0122 \\
\texttt{em} loss & 0.0149 & 0.0627 & 0.0019 & 0.0090 \\
\texttt{greedy} loss & 0.0189 & 0.0767 & 0.0131 & 0.0103 \\
\texttt{sample} loss & 0.0308 & 0.1093 & 0.3978 & 0.0211 \\
\texttt{svd} loss & 0.0203 & 0.0783 & 0.0380 & 0.0143 \\
\hline
\texttt{svd\_w} time (s) & 0.0112 & 0.0115 & 1.0586 & 7.2273 \\
\texttt{adam} time (s) & 0.3883 & 0.3988 & 10.5479 & 63.3477 \\
\texttt{em} time (s) & 0.2356 & 0.3183 & 26.1445 & 206.1457 \\
\texttt{greedy} time (s) & 0.0803 & 0.0895 & 6.4177 & 47.4794 \\
\texttt{sample} time (s) & 0.0672 & 0.0779 & 4.1263 & 36.2301 \\
\texttt{svd} time (s) & 0.0055 & 0.0057 & 0.2831 & 2.2419 \\
\bottomrule
\end{tabular}
\end{table}

\begin{table}[ht]
\caption{Fisher-weighted rank $10$ approximation loss and running time}
\label{tab:experiments-10}
\centering
\begin{tabular}{ l c c c c }
\toprule
Algorithm & \texttt{mnist} & \texttt{fashion\_mnist} & \texttt{smallnorb} & \texttt{colorectal\_histology} \\
\hline
\texttt{svd\_w} loss & 0.0255 & 0.0967 & 0.0198 & 0.0125 \\
\texttt{adam} loss & 0.0245 & 0.0945 & 0.0348 & 0.0145 \\
\texttt{em} loss & 0.0243 & 0.0897 & 0.0202 & 0.0122 \\
\texttt{greedy} loss & 0.0274 & 0.1017 & 0.0391 & 0.0135 \\
\texttt{sample} loss & 0.0410 & 0.1293 & 0.6094 & 0.0240 \\
\texttt{svd} loss & 0.0335 & 0.1094 & 0.0429 & 0.0146 \\
\hline
\texttt{svd\_w} time (s) & 0.0114 & 0.0125 & 0.8855 & 6.1798 \\
\texttt{adam} time (s) & 0.3835 & 0.4199 & 14.9386 & 62.5755 \\
\texttt{em} time (s) & 0.2285 & 0.2776 & 18.3505 & 146.7829 \\
\texttt{greedy} time (s) & 0.0342 & 0.0460 & 3.2980 & 26.5377 \\
\texttt{sample} time (s) & 0.0428 & 0.0421 & 2.3308 & 18.2120 \\
\texttt{svd} time (s) & 0.0049 & 0.0059 & 0.2339 & 1.8703 \\
\bottomrule
\end{tabular}
\end{table}

\begin{table}[ht]
\caption{Fisher-weighted rank $5$ approximation loss and running time}
\label{tab:experiments-5}
\centering
\begin{tabular}{ l c c c c }
\toprule
Algorithm & \texttt{mnist} & \texttt{fashion\_mnist} & \texttt{smallnorb} & \texttt{colorectal\_histology} \\
\hline
\texttt{svd\_w} loss & 0.0364 & 0.1210 & 0.0872 & 0.0161 \\
\texttt{adam} loss & 0.0355 & 0.1173 & 0.1337 & 0.0180 \\
\texttt{em} loss & 0.0349 & 0.1142 & 0.1052 & 0.0155 \\
\texttt{greedy} loss & 0.0368 & 0.1215 & 0.1393 & 0.0168 \\
\texttt{sample} loss & 0.0470 & 0.1417 & 0.7363 & 0.0239 \\
\texttt{svd} loss & 0.0441 & 0.1312 & 0.2979 & 0.0168 \\
\hline
\texttt{svd\_w} time (s) & 0.0094 & 0.0110 & 0.8328 & 6.1427 \\
\texttt{adam} time (s) & 0.3890 & 0.4410 & 14.6393 & 63.4094 \\
\texttt{em} time (s) & 0.1809 & 0.2019 & 18.0931 & 141.1849 \\
\texttt{greedy} time (s) & 0.0185 & 0.0211 & 1.7466 & 15.8465 \\
\texttt{sample} time (s) & 0.0320 & 0.0343 & 1.4726 & 11.4331 \\
\texttt{svd} time (s) & 0.0039 & 0.0050 & 0.2129 & 1.8317 \\
\bottomrule
\end{tabular}
\end{table}

\section{Row norm sampling for weighted low rank approximation}
\label{sec:row-norm-sampling}

We correct an error of a row norm sampling result of \cite{Dai2023}, and further tighten the result and simplify the proof. The algorithm we study is to repeatedly sample rows according to a \emph{row norm sampling} distribution (Definition \ref{def:row-norm-sampling}).

\begin{Definition}[Row norm sampling]
\label{def:row-norm-sampling}
Let $\bfA\in\mathbb R^{n\times d}$. Then, a \emph{row norm sample} from this matrix samples the row $\bfe_i^\top\bfA$ for $i\in[n]$ with probability
\[
    p_i = \frac{\norm{\bfe_i^\top\bfA}_2^2}{\norm{\bfA}_F^2}.
\]
\end{Definition}

We prove the following theorem, which corrects and improves Theorem 3 of \cite{Dai2023}.

\begin{Theorem}
Let $\bfA\in\mathbb R^{n\times d}$ be a rank $d$ matrix and let $\bfW\in\mathbb R^{n\times d}$ be non-negative weights bounded by $1$. Let $\bfA^*\in\mathbb R^{n\times d}$ be a rank $k$ matrix satisfying
\[
    \norm{\bfA-\bfA^*}_{\bfW,F}^2 = \min_{\rank(\bfA')\leq k}\norm{\bfA-\bfA'}_{\bfW,F}^2.
\]
Let $T\subseteq[n]$ be a multiset of $t$ indices sampled according to the distribution of Definition \ref{def:row-norm-sampling}. If $t\geq (2\sqrt{10}+1)^2 \norm{\bfA^*\bfA^-}_F^2 / \eps^2$, then with probability at least $9/10$, there is a matrix $\hat\bfA\in\mathbb R^{n\times d}$ with rows spanned by the rows sampled in $T$ such that
\[
    \norm{\bfA-\hat\bfA}_{\bfW,F}^2 \leq \norm{\bfA-\bfA^*}_{\bfW,F}^2 + \eps \norm{\bfA}_F^2.
\]
\end{Theorem}

We make a couple of remarks about this result before showing the proof.

\begin{Remark}
If $\bfW$ is all ones and $\bfA^* = \bfA_k$ is the optimal rank $k$ approximation in the standard Frobenius norm given by the SVD, then $\norm{\bfA_k\bfA^-}_F^2 = k$ and thus we recover a result of \cite{FKV2004}. We empirically estimate the value of $\norm{\bfA_k\bfA^-}_F^2$ for weighted low rank approximation by treating the best solution we find as the ``true'' solution $\bfA^*$, and we find that this value is $\leq 2k$ on these datasets.
\end{Remark}

\begin{proof}[Proof of Theorem \ref{def:row-norm-sampling}]
Define the random matrix
\[
    \hat\bfA = \frac1t \sum_{i\in T}\frac{\bfA^*\bfA^-\bfe_i\bfe_i^\top\bfA}{p_i}
\]
Note that $\E[\hat\bfA] = \bfA^*$ and that $\hat\bfA$ is in the row span of the rows of $\bfA$ sampled by $T$. Then, the variance of a single sample in $T$ is bounded by
\[
    \sum_{i=1}^n \frac1{p_i} \norm{\bfA^*\bfA^-\bfe_i\bfe_i^\top \bfA}_{F}^2 = \sum_{i=1}^n \frac{\norm{\bfA}_{F}^2}{\norm{\bfe_i^\top\bfA}_{2}^2} \norm{\bfA^*\bfA^-\bfe_i}_2^2 \norm{\bfe_i^\top \bfA}_2^2 = \norm{\bfA^*\bfA^-}_F^2 \norm{\bfA}_{F}^2
\]
so the variance of $\hat\bfA$ is bounded by
\[
    \Var(\hat\bfA) = \E\norm{\bfA^* - \hat\bfA}_F^2 \leq \frac1t \norm{\bfA^*\bfA^-}_F^2 \norm{\bfA}_F^2.
\]
Thus by Markov's inequality,
\[
    \norm{\bfA^*-\hat\bfA}_F^2 \leq \frac{10}t \norm{\bfA^*\bfA^-}_F^2\norm{\bfA}_F^2
\]
with probability at least $9/10$. Then,
\begin{align*}
    \norm{\bfA-\hat\bfA}_{\bfW,F} &\leq \norm{\bfA-\bfA^*}_{\bfW,F} + \norm{\bfA^*-\hat\bfA}_{\bfW,F} && \text{triangle inequality} \\
    &\leq \norm{\bfA-\bfA^*}_{\bfW,F} + \norm{\bfA^*-\hat\bfA}_{F} \\
    &\leq \norm{\bfA-\bfA^*}_{\bfW,F} + \frac{\sqrt{10}}{\sqrt t}\norm{\bfA^*\bfA^-}_F \norm{\bfA}_F.
\end{align*}
Squaring both sides yields
\begin{align*}
    \norm{\bfA-\hat\bfA}_{\bfW,F}^2 &\leq \norm{\bfA-\bfA^*}_{\bfW,F}^2 + 2\norm{\bfA-\bfA^*}_{\bfW,F}\frac{\sqrt{10}}{\sqrt t}\norm{\bfA^*\bfA^-}_F \norm{\bfA}_F + \frac1t \norm{\bfA^*\bfA^-}_F^2 \norm{\bfA}_F^2 \\
    &\leq \norm{\bfA-\bfA^*}_{\bfW,F}^2 + 2\frac{\sqrt{10}}{\sqrt t}\norm{\bfA^*\bfA^-}_F \norm{\bfA}_F^2 + \frac1t \norm{\bfA^*\bfA^-}_F^2 \norm{\bfA}_F^2 \\
    &= \norm{\bfA-\bfA^*}_{\bfW,F}^2 + \parens*{2\frac{\sqrt{10}}{\sqrt t}\norm{\bfA^*\bfA^-}_F + \frac1t \norm{\bfA^*\bfA^-}_F^2 }\norm{\bfA}_F^2 \\
    &\leq \norm{\bfA-\bfA^*}_{\bfW,F}^2 + (2\sqrt{10}+1)\frac{\norm{\bfA^*\bfA^-}_F}{\sqrt t}\norm{\bfA}_F^2 \\
    &\leq \norm{\bfA-\bfA^*}_{\bfW,F}^2 + \eps \norm{\bfA}_F^2.
\end{align*}
\end{proof}

\end{document}